\documentclass[epj]{webofc}
\usepackage[varg]{txfonts}   
\usepackage{braket}
\usepackage{dutchcal}
\usepackage{amsfonts}
\usepackage{amsmath}
\usepackage{subfigure}
\usepackage{graphicx}
\usepackage{hyperref}
\newcommand{\beq}{\begin{equation}}
\newcommand{\eeq}{\end{equation}}
\newcommand{\bea}{\begin{eqnarray}}
\newcommand{\eea}{\end{eqnarray}}
\newcommand{\nn}{\nonumber}

\newcommand{\bs}{\boldsymbol}
\newcommand{\la}{\lambda}
\newcommand{\eps}{\epsilon}
\newcommand{\beqa}{\begin{eqnarray}}
\newcommand{\eeqa}{\end{eqnarray}}
\newcommand{\be}{\begin{equation}}
\newcommand{\ee}{\end{equation}}
\begin{document}
\title{}
\title{Quantum Field Theory and its Anomalies for Topological Matter}
\subtitle{QCD@Work 2022}

\author{\firstname{Claudio } \lastname{Corian\`o}\inst{1}\fnsep\thanks{\email{claudio.coriano@le.infn.it}}, \firstname{Mario} \lastname{Cret\`i}\inst{1}\inst{2}\fnsep\thanks{\email{mario.creti@unisalento.it}}, \firstname{Stefania} \lastname{D'Agostino}\inst{1}\inst{2}\inst{3}\fnsep\thanks{\email{stefania.dagostino@iit.it}}}

\institute{ Dipartimento di Matematica e Fisica, Universit\`a del Salento \\
INFN-Lecce, Via Arnesano, 73100 Lecce, Italy\\
$^2$Center for Biomolecular Nanotechnologies, Istituto Italiano di Tecnologia, Arnesano (LE), Italy. \\
$^3$ CNR Institute of Nanotechnology (NANOTEC), SP Lecce-Monteroni, Lecce, Italy.\\
 }

\abstract{Topology enters in  quantum field theory (qft) in multiple forms: one of the 
most important, in non-abelian gauge theories, being in the identification of the $\theta$ vacuum in QCD. 
A very relevant aspect of this connection is through the phenomenon of chiral and conformal qft anomalies. 
It has been realized that a class of materials, comprising topological insulators and Weyl semimetals, 
also exhibit the phenomenon of anomalies, which are responsible for several exotic phenomena, such as the presence of edge currents,  resilient under perturbations and scattering by impurities.  
Another example comes from the response functions of these materials under thermal and mechanical stresses, that may be performed using correlation functions of stress energy tensors in General Relativity. In this case the conformal anomaly plays an important role. 
We briefly illustrate some salient features of this correspondence, and the effective action describing the long-range interactions  
that may account for such topological effects.
} 

\maketitle

\section{Introduction}
Quantum anomalies have been discovered in particle physics long ago but, more recently, they have found applications in solid state physics as well, in the case of topological materials, such as Dirac and Weyl semimetals \cite{Chernodub:2021nff,Fruchart:2013tza,Arouca:2022psl,Tutschku:2020rjq,Tkachov:2011ir}, with important implications in anomalous transport \cite{Landsteiner:2012kd,Chernodub:2020dqo,Chernodub:2017jcp} in condensed matter and heavy ion collisions \cite{Kharzeev:2022ydx}. Gravity enters the field via the conformal anomaly \cite{Duff:1993wm} (see \cite{Coriano:2020ees} for a review) and the Luttinger relation \cite{Luttinger:1964zz}. 
Dirac semimetals, for example, are 3D crystals whose low-lying excitations are described by the massless Dirac equation. The relativistic description is emergent, in the sense that a linear dispersion relation in the band structure of such materials is associated with the Fermi velocity $v_F$  rather than with the speed of light $c$. 
In a Dirac semimetals the valence and conductance bands touch each other in a single point, called a Dirac point. 
			At such specific point the energy/momentum dispersion relation is linear, similarly to the relativistic case. In Weyl semimetals a Dirac point could also split into a pair of so-called Weyl points, at which the dispersion relation is exactly that of a Weyl Hamiltonian $H^{\pm}_{Weyl}=\pm\vec p\cdot \vec\sigma$. \\
			The response function of such materials may exhibit an effective anomaly if we engineer an axial-vector-vector (AVV) interaction, in which left- and right-handed Weyl modes of the electrons in the material couple with different strengths to an external chiral current, generated via a chiral chemical potential $(\mu_5)$.\\  
The $\bf{AVV}$ anomaly vertex, for example, which is responsible for the chiral magnetic effect, is generated when the material is subjected to a background of a magnetic field as well as to a non-zero chiral chemical potential (the time-like component of the remaining axial current), with the generation of an electric current. Therefore, a combined study of this interaction from the qft and condensed matter theory point of views is crucial in order to understand the phenomenology of such materials. It is expected that the interaction involves an intermediate pseudoscalar state whose properties resemble those of an axion. The analysis of such state, in the long wavelength limit, in which the description of the material can be performed using ordinary chiral/conformal qft actions, raises some interesting issues concerning the type of effective actions which are most suitable for this purpose. Understanding the main features of this interaction and its difference/similarity with respect to the ordinary axion action of the QCD theta-vacuum is the point which we will briefly concentrate upon. \\
All anomalies are associated with effective actions characterised by 
a specific nonlocal structure, in the form of anomaly poles \cite{Giannotti:2008cv}\cite{Armillis:2009im,Coriano:2018zdo,Coriano:2020ees}.  A beautiful unification of such phenomenon is in the case of supersymmetric theories \cite{Coriano:2014gja}, where the conformal and chiral anomalies, together with a supersymmetric anomaly, 
are part of a multiplet. Nonlocal exchanges in the anomaly effective action are identified in all the components of the multiplet. \\
In  qft, chiral anomalies are a relativistic phenomenon connected 
with the presence of Weyl fermions, whose couplings to conserved external currents is different for left (L) and right handed (R) components. A chiral anomaly can be generated at trilinear level - the counting refers to the number of external currents - and anomaly cancelation is necessary if the corresponding interaction violates the Ward identities that result from the original quantum symmetry of the action. \\
If we expand the partition function of a certain theory 
\begin{equation}
\label{one}
\mathcal{Z}(\chi_i) \equiv \int D\psi D\bar{\psi} e^{i\mathcal{S}_0(\psi, \bar{\psi}, B)}
\end{equation}
in the case of the chiral anomaly, the  underlying interaction is that of an $\bf{AVV}$ (axial-vector/vector/vector) anomaly diagram, appearing at $O(e^2)$ in the expansion of \eqref{one}, when $\mathcal{S}_0$ is the fundamental action of axial QED 
\begin{equation}
\label{aqed}
\mathcal{S}_0=\bar{\psi}\gamma^\mu(\partial_\mu -ie A_\mu - i \gamma_5 B_\mu)\psi - m\bar{\psi}\psi
\end{equation}
and $B_\lambda$ is the axial-vector source.
In the presence of propagating axial-vector and vector interactions, in the massless limit, such anomaly interactions have to vanish, by summing over all the fermions species that are coupled to the corresponding gauge fields. In the case of the Standard Model (SM) of the elementary particles, this is realized by summing over each fermion generation of quarks and leptons, though such cancelation may occur, in other models, by a suitable 
choice of the fermion charges, in the form of algebraic costraints  involving several generations. The constraints take the form of linear, quadratic and cubic Diophantine equations. In general, in the case of the AVV diagram, a generic parameterization of this contribution leads to a correlator that violates the ordinary vector and axial vector Ward identities in the form 
 
 \beq
 k_{1\mu} \Delta^{\la\mu\nu}(k_1,k_2) = a_1 \epsilon^{\lambda\nu\alpha\beta} 
k_1^\alpha k_2^\beta,\,\, 
k_{2\nu} \Delta^{\la\mu\nu}(k_1,k_2) = a_2 \epsilon^{\lambda\mu\alpha\beta} k_2^\alpha k_1^\beta, \,\,\,
 k_{\la} \Delta^{\la\mu\nu}(k_1,k_2) = a_3 \epsilon^{\mu\nu\alpha\beta} 
k_1^\alpha k_2^\beta, \nonumber \\
\eeq
where $a_1=-\frac{i}{8 \pi^2} \qquad a_2=-\frac{i}{8 \pi^2} \qquad a_3=-\frac{i}{4 \pi^2}$. In the correlator $\Delta^{\lambda\mu\nu}$, $\mu$ and $\nu$ are vector indices and $\lambda$ an axial-vector one. Notice that $a_1=a_2 $, as expected from the Bose symmetry of the two {\bf V} lines. We have denoted with $k_1$ and $k_2$ the momenta of the two photons. $k$ is the momentum of the axial-vectors source.
It is also well known that the total anomaly 
$a_1+a_2 + a_3 \equiv  a_n$ is regularization scheme independent 
($a_n=-\frac{i}{2 \pi^2}$). A general parameterization involves 6 form factors $(A_i)$ in the expression of the diagram
 
 \begin{eqnarray}
\Delta_0^{\la\mu\nu} &=& A_1 (k_1, k_2) \eps [k_1,\mu,\nu,\la] + A_2 (k_1, k_2)\eps [k_2,\mu,\nu,\la] +
A_3 (k_1, k_2) \eps [k_1,k_2,\mu,\la]{k_1}^{\nu} \nonumber \\
&& +  A_4 (k_1, k_2) \eps [k_1,k_2,\mu,\la]k_2^{\nu}
 + A_5 (k_1, k_2)\eps [k_1,k_2,\nu,\la]k_1^\mu
+ A_6 (k_1, k_2) \eps [k_1,k_2,\nu,\la]k_2^\mu,\nonumber \\
\label{Ros}
\end{eqnarray}
 where $A_1$ and $A_2$ are divergent by power counting. The parameterization will give conserved vector Ward identites if we allow a Chern-Simons (CS) contribution with a parameter $\beta$, introduced by the momentum reparametrization in the loop
 \beq
\Delta^{\la\mu\nu}(\beta,k_1,k_2)= \Delta^{\la\mu\nu}(k_1,k_2) - \frac{i}{4 \pi^2}\beta \epsilon^{\lambda\mu\nu\sigma}\left( k_{1\sigma} - 
k_{2\sigma}\right).
\label{bd}
\eeq
The CS term (the Bardeen counterterm) will affect only $A_1$ and $A_2$, and corresponds to a finite renormalization of the diagram. $A_1$ and $A_2$ will shift by the same counterterm under such reparametrization, while their difference, obviously, will be invariant. As shown in \cite{Armillis:2009sm}, the pole is associated with the difference of these two form factors and is, therefore, unambiguous. The pole shows up once we reformulate \eqref{Ros} in the $W_L/W_T$  (longitudinal/transverse or L/T) decomposition of the same vertex, performed respect to the momentum of the axial-vector current, $k$. The presence of such unique interactions and its contribution to the longitudinal Ward identity of the axial-vector current, is the most direct proof that the 1PI (1-particle irreducible) effective action is affected by these types of massless exchanges. 
The transition from the representation in \eqref{Ros} to the $W_L/W_T$ one, requires repeated use of the Schouten relations, with the generation of massless poles also in the transverse part. The effective action, obtained by integrating out the fermions, in the $W_L/W_T$ for is given by \cite{2009PhLB..682..322A,Armillis:2009sm}
\beqa
\mathcal{S}_{eff}&=&-\int d^4 x \frac{1}{4}F_{\mu\nu}F^{\mu\nu} + a_n \int d^4 x \partial\cdot B\,\Box^{-1}(x,y) F_{\mu\nu}\tilde{F}^{\mu\nu}(y) \nn\\
&& +\int d^4 x \,d^4 y\, d^4 z\, W_T^{\la\mu\nu}(z,x,y) \,B_{\lambda}(z)\, A_{\mu}(x)\, A_{\nu}(y).
\label{int}
\eeqa
where $A_{\mu}$ is the photon. 
In the equation above, beside the kinetic term of the photon field ($F_{\mu\nu}$), the second contribution has a specific 
nonlocal structure with a $1/\Box$ interaction. In momentum space it is identified by the 2-particle cut in the variable $s\equiv k^2$, the virtuality of the axial-vector line. It corresponds to an interaction term of the form
\beq
\label{qq}
\partial\cdot B(x)\,\Box^{-1}(x,y) F_{\mu\nu}\tilde{F}^{\mu\nu}(y),
\eeq 
between the longitudinal component of $B_\lambda$ and the anomaly $(F\tilde F).$  The third contribution is divergenceless and corresponds to the transverse component $W_T$. Its expression can be found in \cite{Armillis:2009sm}. Notice that for on-shell photons, the amplitude reduces just to the anomaly pole-like behaviour given in \eqref{qq} with no extra term. 
A similar separation is introduced in the reconstruction program of the graviton vertices in conformal field theory \cite{Bzowski:2013sza}, since this is also based on a similar decomposition, with the inclusion also of traceless and longitudinal sectors. The generation of such a pole is reminiscent of a collinear interaction with an intermediate fermion-antifermion pair in the loop \cite{Giannotti:2008cv,Armillis:2009pq}, generated by the decay of an off-shell axial-vector current. It can be understood via the Coleman-Norton interpretation of a singularity in a certain diagram, as due to a physical process. Note that in the massive case, the pole of the spectral density turns into the 2-particle cut of a (single) anomaly form factor, whose spectral density can be shown to satisfy a sum rule related to the anomaly. The same behaviour is present for the conformal and superconformal anomaly. The pole reappears from the cut once we perform the conformal (zero fermion mass) limit, and the area under the cut is preserved, being mass-independent.\\
 \section{The nonlocal conformal anomaly effective action}
 Non conserved currents carry significant implications in relativistic hydrodynamics. Indeed, the anomalous charge and energy flows can be described in terms of the chiral/conformal/gravitational anomaly actions, depending on the specific case.  
Certain quantum anomalies, for instance conformal and mixed axial-gravitational anomalies, appear in curved backgrounds 
and involve the energy-momentum tensor \cite{Coriano:2020ees}. In condensed matter systems, these anomalies are realized in an off-equilibrium regime using the Luttinger theory of thermal transport coefficients \cite{Luttinger:1964zz}(see also \cite{Bermond:2022mjo}).  In this theory, a temperature gradient ${\bs \nabla} T$ -- drives a system out of equilibrium,  and it is compensated by a non-uniform gravitational potential $\Phi$, at linear order in its fluctuation
\be
\frac{1}{T} {\bs \nabla} T = - \frac{1}{c^2} {\bs \nabla}  \Phi\,,
\label{eq:Luttinger}
\ee
where $c$ is the speed of light. In the case of a weak gravitational field the gravitational potential $\Phi$,
\be
g_{00} = 1 + \frac{2 \Phi}{c^2}\,,
\label{eq:g00}
\ee
turns Newtonian, with $g_{00}$ the time/time component of the metric. The other components of the metric tensor are unmodified. Luttinger's observation finds its origin in the Tolman--Ehrenfest effect~\cite{Tolman:1930zza,Tolman:1930ona}. 
Tolman concluded that systems in thermal equilibrium under the action of a gravitational field do not have a constant temperature. The local temperature distribution is position dependent, a result that is well-known to relativists and cosmologists and used in several applications in these areas. In other words, they stated that heat has weight, while Luttinger identified a connection between gravitational fields and thermal transport. The temperature depends on the spatial coordinates as
\be
T(x) = T_0 /\sqrt{g_{00}(x)},
\label{eq:TE}
\ee
where $T_0$ denotes a reference temperature, selected at a point with $g_{00}=1$, i.e. at asymptotic distance from a gravitational source. The Luttinger formula~\eqref{eq:Luttinger} may be derived from simple thermodynamic considerations.\\
Once the connection between thermal and mechanical stresses of such materials and gravity is made explicit, the analysis inherits a great deal of the formalism used in the context of general relativity, opening the way to the use of correlation functions in external metric backgrounds in the study of these materials. Renormalization of the matter sector generates a back-reaction of the quantum corrections on the external metric
\cite{Coriano:2022ftl} that is described by an anomaly action, which is, in part, topological.
Its topological nature is evident from the operatorial structure of the equation connecting the trace of the stress energy tensor of a conformal theory with topological invariants such as the Euler-Poincar\`e density, according to the quantum relation  

\be
g_{\mu\nu}\langle T^{\mu\nu}\rangle =b E + b' C^2 + c_1 \Box R 
\label{ann}
\ee
where $E$ and $C^2$ denote the Euler Poincar\`e density (Gauss Bonnet term) and the Weyl tensor squared
\begin{align}
\label{fourd}
E_4&\equiv R_{\mu\nu\alpha\beta}R^{\mu\nu\alpha\beta}-4R_{\mu\nu}R^{\mu\nu}+R^2  \\
( C^{(4)})^2&\equiv R_{\mu\nu\alpha\beta}R^{\mu\nu\alpha\beta}-2R_{\mu\nu}R^{\mu\nu}+\frac{1}{3}R^2, \label{fourd2}
\end{align}
where the first two terms on the rhs of \eqref{ann} originate from a nonlocal effective action.
The $E$ invariant can be generalized to any even dimension, becoming cubic, quartic, and so on, in the Riemann tensor and its contractions
\bea
\label{Ed}
E_d = \frac{ 1}{2^{d/2}}
\delta_{\mu_1 \cdots \mu_d}^{\nu_1 \cdots \nu_d}
{R^{\mu_1 \mu_2}}_{\nu_1 \nu_2} \cdots
{R^{\mu_{d-1} \mu_d}}_{\nu_{d-1} \nu_d} \ ,
\eea
at d=4,6, etc. At $d=2$ this role is taken by the Einstein-Hilbert 
action. These contributions are investigated in dimensional regularization, generating a dilaton in the 
effective theory  that can be removed by a finite renormalization of the Gauss-Bonnet term \cite{Coriano:2022ftl,Coriano:2022knl}.  The anomalous trace of the energy-momentum tensor is known to be generated by the nonlocal action~\cite{Armillis:2009pq,Riegert:1984kt,Mazur:2001aa,Mottola:2006ew}
\beqa
S_{\mathrm{anom}}[g,A] & {=} & \frac{1}{8} \int d^4 x \sqrt{- g(x)}  \int d^4 y \sqrt{- g(y)} 
 \cdot E(x) G^{(4)}(x,y)\times \nonumber \\
 && \times  \left[ 2 b' C^2(y) + b E(y) + 2 c F_{\mu\nu}(y) F^{\mu\nu}(y)\right]\!,
\nonumber
\eeqa
where $G^{(4)}(x, y)$ is the Green function the fourth-order conformally covariant differential operator 
\beq
\Delta_4 = \nabla_\mu \left( \nabla^\mu \nabla^\nu + 2 R^{\mu\nu} - \frac{2}{3} R g^{\mu\nu}\right) \nabla_\nu.
\label{eq:Delta:4}
\eeq
An application of the formalism is provided by a system in a slightly off-equilibrium regime \cite{Chernodub:2019tsx} with a small temperature variation along a 
certain $(x_3)$ direction. We use the Luttinger relation~\eqref{eq:Luttinger} to link the time-independent temperature gradient to the gradient of the gravitational potential $\Phi$. Then, the perturbation of the metric is nonvanishing only for the $h_{00} \equiv 2 \Phi$ component, that in our case is a function of only on one spatial variable, $h_{00}(x) \equiv h_{00}(x_3)$. One finds that the anomalous part of the $TTT$ vertex contributes to the expectation value for the energy-momentum tensor in the form
\begin{align}
\braket{T^{00}}_{TTT}&=\frac{4 b'}{9} \Big[3\Big(\partial_3^2\,\Phi\Big)^2 + 4\Big(\partial_3 \Phi\Big)\,\Big(\partial_3^3 \Phi\Big) + 2 \Phi\partial_3^4\,\Phi\Big],
\label{eq:T00:pre}\\
\braket{T^{11}}_{TTT}&=\braket{T^{22}}_{TTT} = \frac{4 b'}{9}\Big[2\Big(\partial_3 \Phi\Big)\,\Big(\partial_3^3 \Phi\Big) + \Phi\partial_3^4 \Phi\Big].
\label{eq:T11:pre}
\end{align}
 Eqs.~\eqref{eq:T00:pre}--\eqref{eq:T11:pre} show some striking features, one of them being their local expression in terms of the gravitational potential $\Phi \equiv h_{00}/2$, despite the fact that they have been derived from the non-local anomaly action. A second important point is that
  the expectation value of the energy-momentum tensor involves only the anomalous coefficient $b'$. This coefficient, given explicitly in \eqref{ann} for the case of one-species QED, is related to the truly anomalous part of the energy-momentum tensor (the Weyl term $C^2$). Indeed, $E$ is an evanescent counterterm at $d=4$, whose appearance in the Weyl variation of the anomaly action originates by requiring that the same action satisfies the Wess-Zumino consistency condition. Consequently, there is no topological contribution coming from the Euler density to the trace of the energy-momentum tensor at cubic order. These analysis can be extended to higher point functions.
\section{Conclusions}
The study of anomaly effective actions, and in particular their local and nonlocal formulations, finds important applications in the context of topological materials. Here we have briefly described some of these connections. This shows the existence of an 
important interplay between some fundamental areas of high energy physics - anomalies are such - and condensed matter theory. 

\centerline{\bf Acknowledgements} 
We thank M. Chernodub, E. Hankiewicz, M.M. Maglio, E. Mottola and M. Vozmediano for discussions. We thank S. Lionetti and R. Tommasi for collaborating to related analysis. The work of C.C. is partly supported by INFN 
Iniziativa Specifica QFT-HEP.


\begin{thebibliography}{29}

\bibitem{Chernodub:2021nff}
M.N. Chernodub, Y.~Ferreiros, A.G. Grushin, K.~Landsteiner, M.A.H. Vozmediano,
  Phys. Rept. \textbf{977}, 1 (2022), \texttt{2110.05471}

\bibitem{Fruchart:2013tza}
M.~Fruchart, D.~Carpentier, Comptes Rendus Physique \textbf{14}, 779 (2013),
  \texttt{1310.0255}

\bibitem{Arouca:2022psl}
R.~Arouca, A.~Cappelli, T.H. Hansson (2022), \texttt{2204.02158}

\bibitem{Tutschku:2020rjq}
C.~Tutschku, F.S. Nogueira, C.~Northe, J.~van~den Brink, E.M. Hankiewicz, Phys.
  Rev. B \textbf{102}, 205407 (2020), \texttt{2007.11852}

\bibitem{Tkachov:2011ir}
G.~Tkachov, E.M. Hankiewicz, Physica E \textbf{44}, 900 (2012),
  \texttt{1106.1059}

\bibitem{Landsteiner:2012kd}
K.~Landsteiner, E.~Megias, F.~Pena-Benitez, Lect. Notes Phys. \textbf{871}, 433
  (2013), \texttt{1207.5808}

\bibitem{Chernodub:2020dqo}
M.N. Chernodub, E.~Kilincarslan, Phys. Rev. D \textbf{103}, 036019 (2021),
  \texttt{2010.02104}

\bibitem{Chernodub:2017jcp}
M.N. Chernodub, A.~Cortijo, M.A.H. Vozmediano (2017), \texttt{1712.05386}

\bibitem{Kharzeev:2022ydx}
D.E. Kharzeev, \emph{{Chiral magnetic effect in heavy ion collisions and
  beyond}} (2022), \texttt{2204.10903}

\bibitem{Duff:1993wm}
M.J. Duff, Class. Quant. Grav. \textbf{11}, 1387 (1994),
  \texttt{hep-th/9308075}

\bibitem{Coriano:2020ees}
C.~Corian\`o, M.M. Maglio, Phys. Rept. \textbf{952}, 2198 (2022),
  \texttt{2005.06873}

\bibitem{Luttinger:1964zz}
J.M. Luttinger, Phys. Rev. \textbf{135}, A1505 (1964)

\bibitem{Giannotti:2008cv}
M.~Giannotti, E.~Mottola, Phys. Rev. \textbf{D79}, 045014 (2009),
  \texttt{0812.0351}

\bibitem{Armillis:2009im}
R.~Armillis, C.~Corian\`o, L.~Delle~Rose, Phys. Lett. \textbf{B682}, 322
  (2009), \texttt{0909.4522}

\bibitem{Coriano:2018zdo}
C.~Corian\`o, M.M. Maglio, Phys. Lett. \textbf{B781}, 283 (2018),
  \texttt{1802.01501}

\bibitem{Coriano:2014gja}
C.~Corian\`o, A.~Costantini, L.~Delle~Rose, M.~Serino, JHEP \textbf{06}, 136
  (2014), \texttt{1402.6369}

\bibitem{Armillis:2009sm}
R.~Armillis, C.~Corian\`o, L.~Delle~Rose, M.~Guzzi, JHEP \textbf{12}, 029
  (2009), \texttt{0905.0865}

\bibitem{2009PhLB..682..322A}
R.~{Armillis}, C.~{Corian{\`o}}, L.~{Delle Rose}, Physics Letters B
  \textbf{682}, 322 (2009), \texttt{0909.4522}

\bibitem{Bzowski:2013sza}
A.~Bzowski, P.~McFadden, K.~Skenderis, JHEP \textbf{03}, 111 (2014),
  \texttt{1304.7760}

\bibitem{Armillis:2009pq}
R.~Armillis, C.~Corian\`{o}, L.~Delle~Rose, Phys. Rev. \textbf{D81}, 085001
  (2010), \texttt{0910.3381}

\bibitem{Bermond:2022mjo}
B.~Bermond, M.~Chernodub, A.G. Grushin, D.~Carpentier (2022),
  \texttt{2206.08784}

\bibitem{Tolman:1930zza}
R.C. Tolman, Phys. Rev. \textbf{35}, 904 (1930)

\bibitem{Tolman:1930ona}
R.~Tolman, P.~Ehrenfest, Phys. Rev. \textbf{36}, 1791 (1930)

\bibitem{Coriano:2022ftl}
C.~Corian\`o, M.M. Maglio, D.~Theofilopoulos (2022), \texttt{2203.04213}

\bibitem{Coriano:2022knl}
C.~Corian\`o, M.M. Maglio, Phys. Lett. B \textbf{828}, 137020 (2022),
  \texttt{2201.07515}

\bibitem{Riegert:1984kt}
R.J. Riegert, Phys. Lett. \textbf{134B}, 56 (1984)

\bibitem{Mazur:2001aa}
P.O. Mazur, E.~Mottola, Phys.Rev. \textbf{D64}, 104022 (2001),
  \texttt{hep-th/0106151}

\bibitem{Mottola:2006ew}
E.~Mottola, R.~Vaulin, Phys. Rev. \textbf{D74}, 064004 (2006),
  \texttt{gr-qc/0604051}

\bibitem{Chernodub:2019tsx}
M.N. Chernodub, C.~Corian\`o, M.M. Maglio, Phys. Lett. \textbf{B802}, 135236
  (2020), \texttt{1910.13727}

\end{thebibliography}

\end{document}